# An information circuit driven by a skyrmion Maxwell's demon

Soma MIKI[1,2] *†, Yoshishige SUZUKI[2–8*], Ryo ISHIKAWA[4,5,9,10], Yuki HIBINO[6], Shunya MORISHITA[9], Hiroto IMANISHI[2], Keigo ADACHI[2], Yoichi SHIOTA[11,12], Takayuki NOZAKI[6], Minori GOTO[5,13], Hikaru NOMURA[5,14], Shigemi MIZUKAMI[1,8], and Eiiti TAMURA[2,15]

1. *WPI Advanced Institute for Materials Research (AIMR), Tohoku University, Sendai, 980-8577, Japan*
2. *Graduate School of Engineering Science, The University of Osaka, Toyonaka, 560-8531, Japan*
3. *Nano-materials Micro-devices Research Center, Osaka Institute of Technology, Osaka, 535-8585, Japan*
4. *Spintronics Research Network Division, Institute for Open and Transdisciplinary Research Initiatives, The University of Osaka, Suita, 565-0871, Japan*
5. *Center for Spintronics Research Network (CSRN), Graduate School of Engineering Science, The University of Osaka, Toyonaka, 560-8531, Japan*
6. *National Institute of Advanced Industrial Science and Technology (AIST), Research Institute for Hybrid Functional Integration, Tsukuba, Ibaraki 305-8568, Japan*
7. *Center for the Promotion of Advanced Interdisciplinary Research, The University of Osaka, Toyonaka, 560-8531, Japan*
8. *Center for Science and Innovation in Spintronics, Core Research Cluster, Tohoku University, Sendai 980-8577, Japan*
9. *Graduate School of Engineering, The University of Osaka, Suita, 565-0871, Japan*
10. *ULVAC– The University of Osaka Joint Research Laboratory for Future Technology, The University of Osaka, Suita, 565-0871, Japan*
11. *Institute for Chemical Research, Kyoto University, Uji, 611-0011, Japan*
12. *Center for Spintronics Research Network (CSRN), Institute for Chemical Research, Kyoto University, Uji, 611-0011, Japan*
13. *Faculty of Advanced Engineering, Tokyo University of Science, Katsushika, Tokyo 125-8585, Japan*
14. *International Center for Synchrotron Radiation Innovation Smart, Tohoku University, Sendai, 980-8579, Japan*
15. *Department of Electronic Science and Engineering, Kyoto University, Kyoto 615-8510, Japan*

* These authors contributed equally.
†Corresponding Author: Soma Miki (soma.miki.d8@tohoku.ac.jp)

# Abstract

**At the nanoscale, observation and feedback may transform thermal fluctuations into valuable resources. Nevertheless, a framework for experimentation with the resulting probability currents has yet to be firmly established. Here, we present an information heat engine incorporating two interacting magnetic skyrmions in adjacent potential wells. Feedback from a skyrmion in Brownian motion at room temperature to the other yields a Maxwell's demon and a circulating probability current in their joint state space. A linear relationship is confirmed between the probability current and the information affinity, the thermodynamic conjugate value, within certain response range and conditions. We identify the slope as the "information conductance" analogous to electrical conductance. The information conductance is modified by skyrmion interactions and scattering, while Kirchhoff's laws remain valid. These results establish a circuit representation of the state space of an information heat engine with potential applications to computing devices, biological systems and chemical reactions.**

# Skyrmion information circuit

Reducing the energy cost of computation—a pressing challenge in the age of AI—requires a framework for circuit design based on information thermodynamics [1]. Information thermodynamics evolved from attempts to resolve the 'Maxwell's demon' paradox—which appears to contradict the second law of thermodynamics by extracting work from thermal fluctuations [2–10], and has now come to demonstrate the fundamental equivalence of information and energy [11]. According to information thermodynamics, an irreversible 1-bit operation requires, in principle, an energy cost of at least $k_{\mathrm{B}}T \ln 2$ [12,13]. Here, $k_{\mathrm{B}} \approx 1.38 \times 10^{-23}$ [J/K] is the Boltzmann constant, and $T$ [K] is the temperature. However, modern computers dissipate $10^3$ to $10^6$ times the energy of this fundamental limit [14]. Reversible computing [15] and Brownian computing [16] have been proposed as ways to overcome this inefficiency. However, practical implementation of the Brownian computing, for example, requires the incorporation of Maxwell's demon (Fig. 1 (a)) into the computational architecture to ensure a scalability of computation time [2, 17–21].

The Horowitz–Esposito (HE) information flow $\dot{I}$ [J/(K·s)] provides a powerful framework for analysing non-equilibrium bipartite systems with feedback [22]. The HE information flow is expressed as the product of the information affinity $F_{\mathrm{S}}^{\mathrm{Info}}$ [J/K] and the probability current $J(C)$ [1/s] flowing through the information circuit in the state space (Fig. 1 (b)) and represents the additional entropy production rate occurring within the subsystem. Including HE information-flow term generalises the second law of thermodynamics at the subsystem level, allowing information to be converted into useful work. Furthermore, in a subsystem that possesses no affinity other than Information affinity, $J(C)$ is a quantity thermodynamically conjugate to $F_{\mathrm{S}}^{\mathrm{Info}}$. Consequently, it is predicted that a relationship analogous to Ohm's law laws holds between $J(C)$ and $F_{\mathrm{S}}^{\mathrm{Info}}$ [23]. However, experimental verification has been difficult to date, as it requires the detection of the microscopic states of a many-body system, real-time feedback, and the observation of sufficiently large number of stochastic events.

Here, we overcame these difficulties by utilising magnetic skyrmions [24,25]. The magnetic skyrmions used in this study are topologically protected, particle-like magnetic structures with a diameter of approximately 1–2 μm, formed within a ferromagnetic thin film. Magnetic skyrmions are a unique condensed matter system that exhibits Brownian motion at room temperature without net material transport [26–29], and they exert repulsive forces on one another [30,31]. Furthermore, their generation, annihilation and motion can be electrically controlled by applying voltage or current [32–38], and their potential landscape can be designed using lithography [39]. Observations of mutual information in a state of thermal equilibrium have shown that skyrmion systems can serve as model systems for information circuits [31].

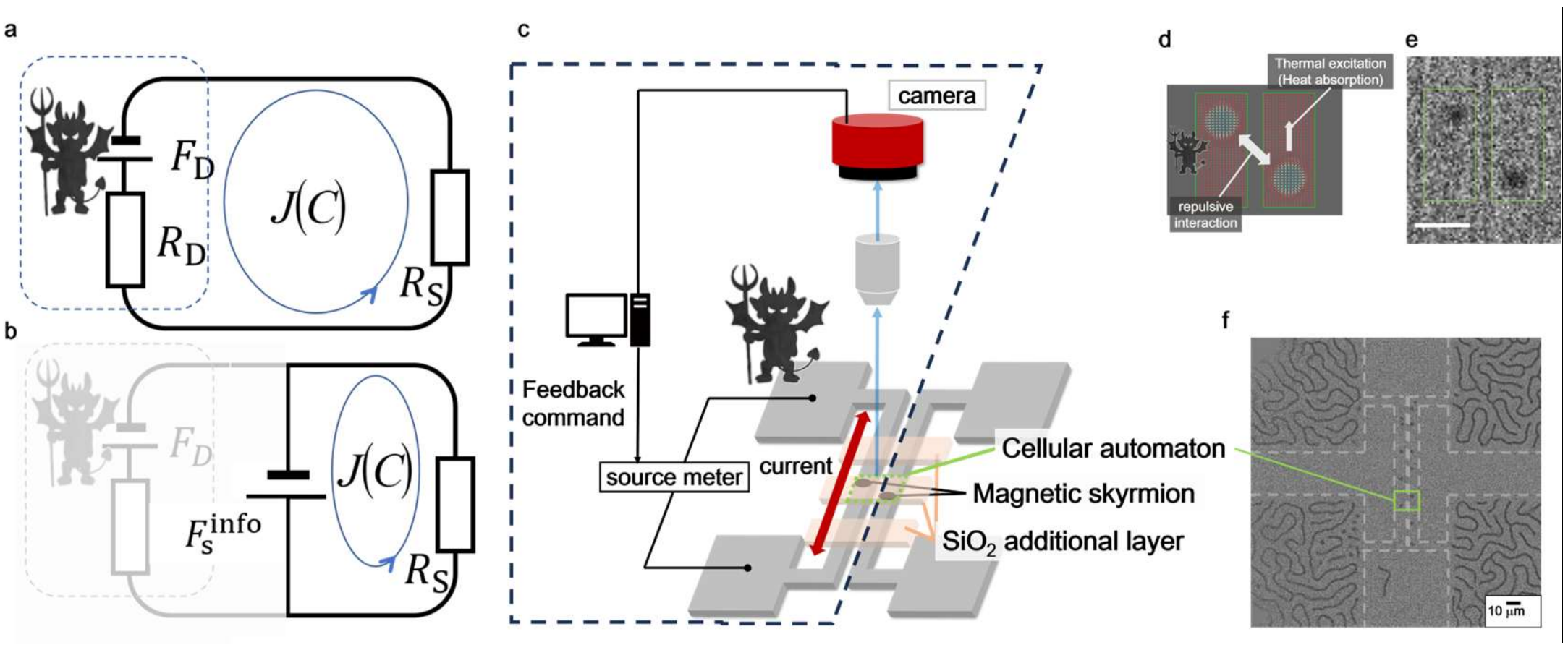

Figure 1 | Experimental realisation of Maxwell's demon using magnetic skyrmions.

**a**, Conceptual circuit of the information engine in a state space. Maxwell's demon is represented by a battery with an internal resistance, which produce a loop affinity and a circulating probability current $J(C)$.

**b**, The circuit is separated to "demon" and "system". Then, Maxwell's demon and the internal resistance are replaced by an information affinity, $F_S^{\text{Info}}$. Here, $\dot{I} = -J(C)F_S^{\text{Info}}$ is the HE information flow expressing a negative entropy flow provided by the demon to the system (right-hand side circuit).

**c**, Two neighbouring nanowires contain elongated strip-shaped potential wells, each confining a single magnetic skyrmion. The position of the skyrmion on the right (the system) is monitored optically, and this information is used in real time to apply an electrical current to the skyrmion on the left (the demon).

**d**, Schematic diagram of skyrmions, each confined within one of two adjacent rectangular potential wells.

**e**, Magneto-optical microscope image of the skyrmions, each confined within one of two adjacent rectangular potential wells

**f**, Magneto-optical microscope image of the device. The green square: a single skyrmion cellular automaton; grey dashed lines: device edge. The two dark spots inside the green square are the magnetic skyrmions.

In this study, we have realised an information engine comprising two magnetic skyrmions undergoing Brownian motion at room temperature. We have demonstrated that the probability current generated therein is proportional, over a relatively wide range, to the non-equilibrium component of the Information affinity generated by Maxwell's demon. Furthermore, we have shown that, in actual information circuits, Ohm's law for information affinity requires slight modifications to account for the interactions between information carriers and the scattering of the probability current.

# Information feedback dependences

Figure 1c shows a schematic diagram of the experimental setup, and Figure 1d shows a magneto-optical micrograph of the device (see 'Overview of Methods'). Within the orange region in Figure 1c, which represents the additional $SiO_2$ layer, the skyrmion on the left-hand magnetic wire is referred to as the 'demon', whilst that on the right-hand magnetic wire is referred to as the 'system'. Each skyrmion is confined within an adjacent rectangular potential well. As shown in Figure 2a, its position assumes two states: '0' at the top and '1' at the bottom. The combination of these two positions allows the dynamics of the entire system to be described as transitions between four states: $n$ = I, II, III and IV. Under conditions where measurements are sufficiently rapid, the demon and the system do not transition simultaneously, forming a bipartite system. Consequently, the transition network reduces to a single closed loop

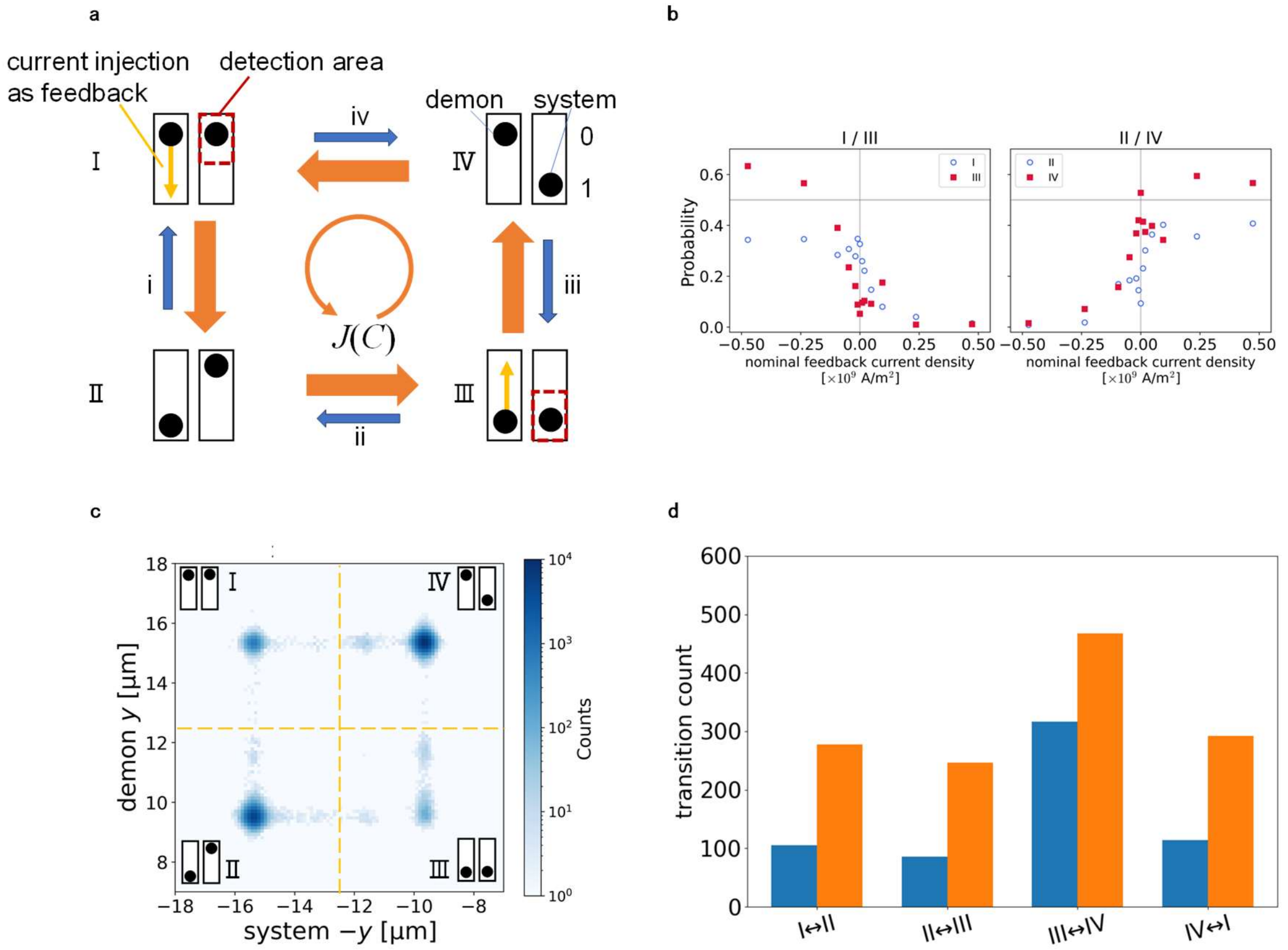

Figure 2 | Four-state representation and transition dynamics of the skyrmion information engine.

**a**, When the two skyrmions never transition simultaneously, the dynamics form a bipartite system. The four states (vertices) connected by permitted transitions (edges) form a single closed loop. When the nominal current is positive, demon transitions from vertex I to II and from III to IV (orange arrows) are favoured. When it is negative, the opposite transitions (blue arrows) occur. Feedback control generates a unidirectional probability current circulating around the loop.

**b**, Probability of each vertex as a function of electric current density under feedback control.

**c**, 2D histogram of the system and demon positions under a feedback current density of $0.094 \times 10^9$ A/m².

**d**, Number of observed transitions between neighbouring states. Although the transition frequencies differ from edge to edge, the difference between forward and backward transitions is identical for all edges, demonstrating the conservation of probability.

$C$ consisting of four vertices (states) and the edges i, ii, iii and iv connecting them.

It is known that even more complex circuits can be similarly decomposed into a superposition of multiple closed loops [22, 23, 40]. Consequently, it is possible to generalise the results of this paper to complex information circuits. Let $p_n$ be the probability that the skyrmion joint system exists at vertex $n$, and let $W_{n\to n+1}$ [$s^{-1}$] be the transition rate from vertex $n$ to $n$+1; then the probability current $J_n$ along each edge is given by:

$$J_n = W_{n\to n+1}\, p_n - W_{n+1\to n}\, p_{n+1}. \tag{1}$$

In this experiment, the position of the system's skyrmion was continuously monitored using a magneto-optical microscope, and a current was applied to the stripe containing the demon skyrmion to realise feedback. When the nominal current is positive, transitions of the demon skyrmion from vertex I to II, and from III to IV, are promoted. When the nominal current is negative, the reverse transitions are promoted. In this feedback cycle, no external force is directly applied to the system's skyrmion itself.

Figure 2b shows how the probability of a skyrmion being present at each of the four vertices depends on the nominal feedback current density. As the positive nominal current density increases, the probabilities at vertices II and IV increase, whilst those at vertices I and III decrease. Figure 2c shows a two-dimensional mapping of the probability distributions of the positions of the system's skyrmion and demon skyrmion at a feedback current density of 0.094 × $10^9$ A/m$^2$. This figure corresponds to the joint probability distribution of their positions. The presence of four clearly distinguishable states (vertices) is evident. The thin lines connecting the four vertices vertically and horizontally appear because skyrmions in the process of transition were observed. On the other hand, no lines are visible in the diagonal directions (connecting I and III, and II and IV), indicating that this system is a bipartite system. Along the lines connecting I and II, and III and IV, there are regions where the probability density is locally slightly higher; these are thought to be weak trap sites. However, no trap sites are observed near the threshold, and the system can be well separated to the four discrete states. Vertices II and IV (diagonal configuration) are energetically stable due to the repulsive interaction between skyrmions, and a positive nominal feedback current increases the probability of these vertices. On the other hand, vertices I and III (parallel configuration) are high-energy states, excited by thermal excitation of skyrmions or a negative nominal feedback current. In Figure 2b, when a large positive nominal current is applied, the probabilities of vertices II and IV should ideally converge to 0.5, respectively. In fact, in this device, these probabilities are approximately 0.4 and 0.6, respectively, indicating this trend. The fact that both probabilities do not exactly reach 0.5 is thought to be due to slight inhomogeneities in the depth of the potential experienced by the skyrmions at the four vertices.

Figure 2d shows the number of skyrmion transitions measured at a feedback current density of 0.094 × $10^9$ A/m$^2$. The blue and orange bars represent clockwise and anti-clockwise transitions, respectively, in the state space shown in Figure 2a. In the absence of feedback control, bidirectional transitions along each edge are in equilibrium, and no net stochastic information flow arises (not shown). Conversely, when the aforementioned automatic feedback procedure (Figure 2a) is introduced, a net probability current $J(C)$ arises along the closed loop $C$, circulating steadily in the $n \to n+1$ direction. This net circulation is directly evidenced by the fact that the number of orange transitions exceeds that of blue ones. The probability current corresponding to the difference in the number of orange and blue transitions is constant along all edges, experimentally demonstrating that Kirchhoff's current law holds at each vertex.

As shown in Equation (1), the probability current arises from transition rates and probability biases. These biases are represented by the bare affinity $F^{\mathrm{Bare}}$ and the information affinity $F^{\mathrm{Info}}$, defined for each edge as follows.

$$\begin{cases} F_n^{\mathrm{Bare}} \equiv k_{\mathrm{B}} ln[W_{n\to n+1}/W_{n+1\to n}] \\ F_n^{\mathrm{Info}} \equiv k_{\mathrm{B}} ln[p_n/p_{n+1}] \end{cases}. \tag{2}$$

Affinity corresponds to voltage in an electrical circuit. The product of affinity and the probability current yields the entropy generation rate $\dot{\sigma}$. In this case, the inclusion of information affinity ensures that the second law of thermodynamics is satisfied in the system [8, 41–43].

$$\dot{\sigma}_{\mathrm{S}} = J(C)\left(F_{\mathrm{S}}^{\mathrm{Bare}} + F_{\mathrm{S}}^{\mathrm{Info}}\right) \geq 0. \tag{3}$$

Here, $\dot{\sigma}_{\mathrm{S}}$ is the entropy generation rate in the system and $F_{\mathrm{S}}^{\mathrm{Bare}} \equiv F_{\mathrm{ii}}^{\mathrm{Bare}} + F_{\mathrm{iv}}^{\mathrm{Bare}}$, $F_{\mathrm{S}}^{\mathrm{Info}} \equiv F_{\mathrm{ii}}^{\mathrm{Info}} + F_{\mathrm{iv}}^{\mathrm{Info}}$. It should be noted that the law concerning entropy in a subsystem [Equation (3)] holds only when information affinity is introduced.

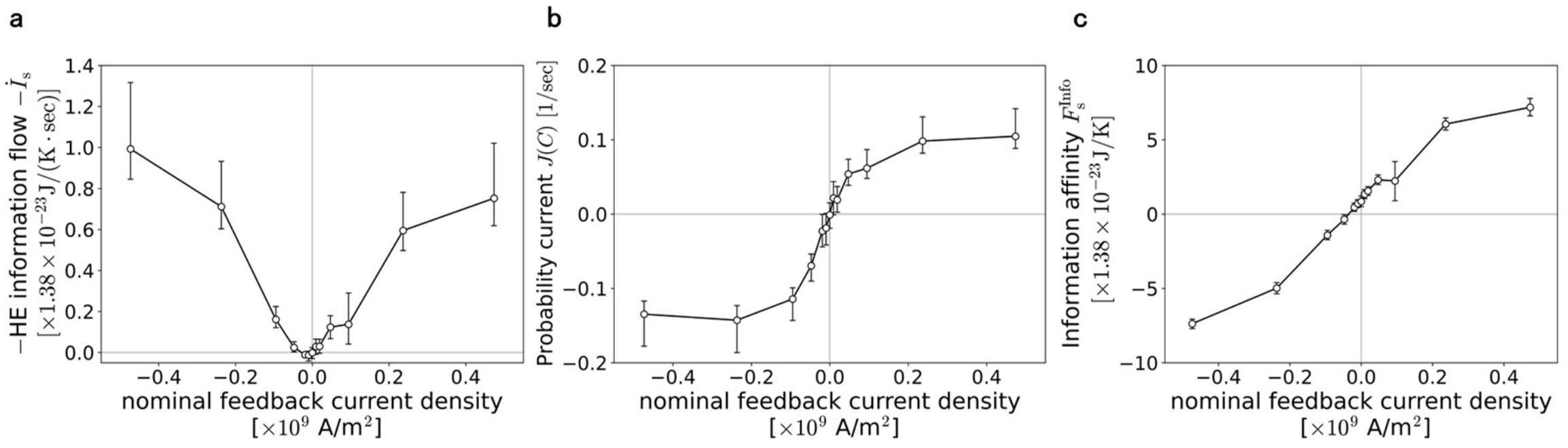

Figure 3 | **a** HE information flow, **b** Probability current, **c** Information affinity: dependence on nominal feedback current. The data points represent experimental results. Solid lines: connecting the data points and serving as a guide for the eye. Error bars: the $3\sigma$ -equivalent statistical uncertainty, defined by the 0.135th and 99.865th percentiles of the bootstrap distribution, together with the systematic uncertainty due to sample degradation.

The HE information flow is given by the following equation.

$$\dot{I}_S \equiv -J(C)F_{\mathrm{S}}^{\mathrm{Info}}. \tag{4}$$

When the demon receives information from the system via feedback, $\dot{I}_S$ becomes negative, and the system can convert information into work.

Figure 3 shows the dependence of (a) HE information flow $\dot{I}_{\mathrm{S}}$, (b) probability current $J(C)$, and (c) information affinity $F_S^{\mathrm{Info}}$ on the nominal feedback current. The HE information flow was calculated from the initial gradients of mutual information as functions of the time delay with the subsystems (Supplementary Information S1). To the best of our knowledge, these are the first experimental observations of information flow, information affinity and probability current in a magnetic skyrmion system [31]. As shown in Figure 2b, as the nominal feedback current increases, the probability distribution reaches saturation. Consequently, the information affinity and probability current also saturate. When the nominal feedback current is zero—that is, when no feedback control is applied, the probability current is zero, and the system reaches thermal equilibrium. However, even when the feedback current is zero, the information affinity $F_{\mathrm{S}}^{\mathrm{Info,eq}}$ in the equilibrium state retains a positive value. This discrepancy stems from the interaction $\varepsilon^{Int}$ between the two skyrmions, which gives rise to a negative 'bare' affinity: $-F_{\mathrm{S}}^{\mathrm{Int}} = -\frac{2k_{\mathrm{B}}\varepsilon^{\mathrm{Int}}}{k_{\mathrm{B}}T} < 0$. From Equation (1), $J(C) = 0$ implies that $F_{\mathrm{S}}^{\mathrm{Info,eq}} = F_{\mathrm{S}}^{\mathrm{Int}} > 0$.

# Information conductance

In this experiment, the feedback current does not exert a direct force on the system. Consequently, the 'bare' affinity is thought to arise solely from the skyrmion–skyrmion interaction. If this is correct, $F_{\mathrm{S}}^{\mathrm{Bare}} + F_{\mathrm{S}}^{\mathrm{Info}}$ in Equation (3) can be replaced by $F_{\mathrm{S}}^{\mathrm{Info}} - F_{\mathrm{S}}^{\mathrm{Info,eq}}$, and the system's entropy generation rate is expressed as follows.

$$\begin{cases} \dot{\sigma}_{\mathrm{S}} = J(C)F_{\mathrm{S}}^{\mathrm{Info,ne}} \geq 0 \\ F_{\mathrm{S}}^{\mathrm{Info,ne}} \equiv F_{\mathrm{S}}^{\mathrm{Info}} - F_{\mathrm{S}}^{\mathrm{Info,eq}} \end{cases}. \tag{5}$$

Here, $F_{\mathrm{S}}^{\mathrm{Info,ne}}$ represents the non-equilibrium component of the information affinity and is zero in the equilibrium state. If Equation (5) holds, $J(C)$ is thermodynamically conjugate to the $F_{\mathrm{S}}^{\mathrm{Info,ne}}$ and can therefore be referred to as an 'information current'.

The following linear response relationship is predicted between thermodynamically conjugate quantities near thermal equilibrium [23,44] (see Supplementary Information S5).

$$J(C) = L_{\mathrm{S}}^{\mathrm{Info}} F_{\mathrm{S}}^{\mathrm{Info,ne}}. \tag{6}$$

Here, $L_{\mathrm{S}}^{\mathrm{Info}}$ [K/(J·s)] is the Onsager coefficient corresponding to electrical conductance in Ohm's law [23], which we term 'information conductance'.

In particular, when edges i and iii, and edges ii and iv are symmetric, respectively, and the system is Markovian, the information conductance can be analytically derived from the master equation as follows (Supplementary Information S3–S5).

$$\begin{cases} L_{\mathrm{S}}^{\mathrm{Info}} = \dfrac{1}{R_{\mathrm{S}}} \\ R_{\mathrm{S}} \equiv \dfrac{4k_{\mathrm{B}}}{W_{\mathrm{II}\to\mathrm{III}}^{\mathrm{eq}}} + \dfrac{4k_{\mathrm{B}}}{W_{\mathrm{III}\to\mathrm{II}}^{\mathrm{eq}}} \end{cases} . \#(7)$$

Here, $R_{\mathrm{S}}$ is the resistance of the system, and $W_{\mathrm{II}\to\mathrm{III}}^{\mathrm{eq}}\, W_{\mathrm{III}\to\mathrm{II}}^{\mathrm{eq}}$ is the transition rate in the absence of feedback. The factor four arises from the normalisation of the state probabilities in the state space; the presence of this coefficient represents a characteristic difference between electrical circuits and state-space circuits (see Supplementary Information S3 and S4), i.e. resistance depends on the size of the state-space.

The circles in Figure 4a indicate the relationship between $J(C)$ and the information affinity $F_{\mathrm{S}}^{\mathrm{Info}}$ obtained from the experiments. The solid line represents a linear fit to the experimental data where the information affinity lies within the range of ±0.3 $k_{\mathrm{B}}$. The value of $\varepsilon^{\mathrm{Int}}$, determined by comparing this fitted line with Equations (5) and (6), is 0.47 $k_{\mathrm{B}}T$, which agrees within the margin of error with $\varepsilon^{\mathrm{Int}} = (0.50 \pm 0.05)k_{\mathrm{B}}T$, obtained from the measured current dependence of the Information affinity shown in Figure 3c.

In contrast, the dotted line in Figure 4a represents the linear relationship between $F_{\mathrm{S}}^{\mathrm{Info}}$ and $J(C)$, calculated by substituting the experimentally obtained parameters into Equations (6) and (7). As can be seen, the line does not reproduce the experimental results. A detailed analysis of the experimental data reveals that the bare affinity of the system skyrmion depends on the feedback current, which may be the cause of the discrepancy. This feedback-current dependence of the system skyrmion can be attributed to an indirect force exerted on the system skyrmion via the 'demon', even though the feedback current does not directly act on the system skyrmion. This indirect interaction leads to scattering of the probability current (Supplementary Information S3, S4). If we denote the system's affinity arising from scattering as $-F_{\mathrm{S}}^{\mathrm{Scattering}}$, Equation (6) should be modified as follows.

$$J(C) = L_{\mathrm{S}}^{\mathrm{Info}}\left(F_{\mathrm{S}}^{\mathrm{Info,ne}} - F_{\mathrm{S}}^{\mathrm{Scattering}}\right). \quad (8)$$

Figures 4c and 4d show the equivalent circuit for this case. The arrows on the battery in the figures indicate that the voltage varies with the feedback strength. Here, considering that within the linear range, $J(C), F_{\mathrm{S}}^{\mathrm{Info,ne}}$ and $F_{\mathrm{S}}^{\mathrm{Scattering}}$ are proportional to the demon's affinity $F_{\mathrm{D}}^{\mathrm{Pump}}$, the effect of scattering can be expressed as a correction to Equations (6) and (7).

$$\begin{cases} J(C) = L_{\mathrm{S}}^{\mathrm{Info}*} F_{\mathrm{S}}^{\mathrm{Info,ne}} \\ L_{\mathrm{S}}^{\mathrm{Info}*} \equiv \dfrac{1-\alpha}{R_{\mathrm{S}} + \alpha R_D} \\ R_{\mathrm{D}} \equiv \dfrac{4k_{\mathrm{B}}}{W_{\mathrm{I}\to\mathrm{II}}^{\mathrm{eq}}} + \dfrac{4k_{\mathrm{B}}}{W_{\mathrm{II}\to\mathrm{I}}^{\mathrm{eq}}} \\ \alpha \equiv \dfrac{F_{\mathrm{S}}^{\mathrm{Scattering}}}{F_{\mathrm{D}}^{\mathrm{Pump}}} \end{cases} . \quad (9)$$

The experimental information conductance, obtained from the slope of the fitting line, is $0.047/k_{\mathrm{B}}$ [K/(J·s)]. The corrected information conductance $L_{\mathrm{S}}^{\mathrm{Info}*}$, calculated from Equation (9) using parameters derived from the measured transition rate, is (0.055±0.025)/ $k_{\mathrm{B}}$ [K/(J·s)] (Supplementary Information S3–S5), and agrees within the margin of error. In this experiment, the value of α is 0.33. Thus, it has been shown that, just as with the relationship between current and voltage in an electrical circuit, a linear relationship holds within a certain range between the probability current and the information affinity, and that this relationship can be explained by correctly incorporating interaction and scattering affinities.

At higher affinities $\left|F_{\mathrm{Sys}}^{\mathrm{Info}}\right| > 5k_{\mathrm{B}}$, the data deviate from a linear response. This deviation may arise because the probability approaches 0.5 and the probability current begins to saturate.

# Kirchhoff's law

First of all, a law similar to the Kirchhoff's current law in electric circuit always holds, due to the conservation of probability. A law corresponding to Kirchhoff's voltage law dictates that the sum of affinities and affinity drops along a closed loop must be zero, i.e. $\sum_{\mathrm{Loop}} F_n - J(\mathrm{C})R_n = 0$ [45]. It can be readily shown that, for affinities satisfying detailed balance, the sum along a closed loop is zero. That is, the sum of the Information affinity and the affinity arising from skyrmion interactions is zero. Maxwell's demon affinity $F_{\mathrm{D}}^{\mathrm{Pump}}$ and the scattering affinity $F_{\mathrm{S}}^{\mathrm{Scattering}}$, which break detailed balance, cancels with affinity drop occurring in the system's resistance and demon's internal resistance. Consequently, Kirchhoff's voltage law is fulfilled both in the total system loop and the system sub-circuit [22,23].

$$\begin{cases} \sum\limits_{n \in Loop} F_n - J(\mathrm{C})\, R_n = F_{\mathrm{D}}^{\mathrm{Pump}} - F_{\mathrm{S}}^{\mathrm{Scattering}} - J(\mathrm{C})(R_S + R_D) = 0 \\ \sum\limits_{n \in System} F_n - J(\mathrm{C})\, R_n = F_{\mathrm{S}}^{\mathrm{Info}} - F_{\mathrm{S}}^{\mathrm{Int}} - F_{\mathrm{S}}^{\mathrm{Scattering}} - J(\mathrm{C}) R_S = 0 \end{cases} .. \quad (10)$$

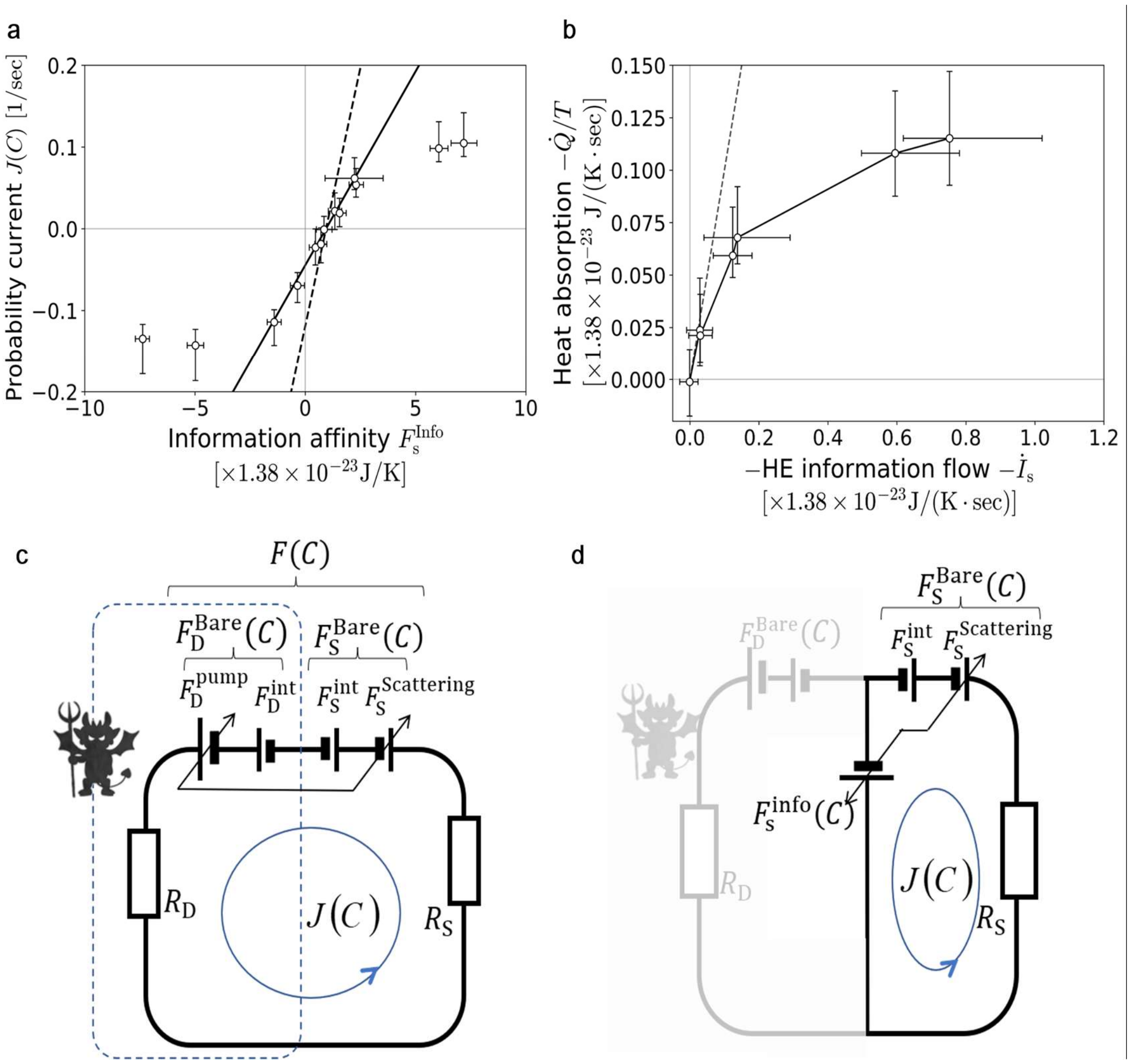

Figure 4 | Linear relation between the probability current and information affinity and heat absorption driven by information feedback.

**a,** Probability current as a function of information affinity: circles, experimental data; solid line; linear fits excluding the two endpoints; dashed line, corresponding to the probability current obtained using Equations (6) and (7); Error bars: the$3\sigma$ -equivalent statistical uncertainty, defined by the 0.135th and 99.865th percentiles of the bootstrap distribution, together with the systematic uncertainty due to sample degradation. **b,** Estimated heat absorbed by the system as a function of the Horowitz–Esposito (HE) information flow: circles, experimental values; dashed line, ideal heat absorption expected for perfect information-to-energy conversion; solid line, a guideline connecting the experimental data points. Error bars: the $3\sigma$ -equivalent statistical uncertainty, defined by the 0.135th and 99.865th percentiles of the bootstrap distribution, together with the systematic uncertainty due to sample degradation. **c,** An equivalent circuit in the state space of the actual experimental setup. The arrows on the batteries signify that their magnitudes vary in correlation with the strength of the feedback. **d,** Equivalent circuit of the system.

# Maxwell's demon

In this system, the HE information flow implies that the system absorbs heat from the environment. We therefore estimated the absorbed heat as $\dot{Q} = 2J(C)\varepsilon^{\mathrm{Int}}$ and plotted it as a function of $-\dot{I}_{\mathrm{S}}$ in Figure 4b. For positive nominal currents, the absorbed heat is positive, indicating that the skyrmion system can function as a refrigerator. This conversion of information into energy constitutes thermodynamic evidence for Maxwell's demon and demonstrates its realisation in a feedback-controlled skyrmion system.

The information–to–energy conversion efficiency, estimated from data points where the HE information flow lies within the range of 0 to 0.4 $k_{\mathrm{B}}$ [J/(K·s)], reaches 23.6 per cent. In principle, as the HE information flow approaches zero, the circuit approaches the reversible limit, and the efficiency approaches 100 per cent (dotted line). On the other hand, when feedback via a negative current is applied, heat is generated.

# Energy efficient information circuits

In this experiment, we utilised the motion of skyrmions driven by an electric current. However, as driving skyrmions with an electric current results in energy loss due to Joule heating, devices utilising electric current remain far from achieving ultra-low power consumption. To minimise power consumption, it is necessary to realise Maxwell's demon by utilising voltage-controlled skyrmions [32–35]. Indeed, we have confirmed through simulation that Maxwell's demon and Ohm's law hold true under voltage control (Supplementary Information S7).

This experiment demonstrates that a relationship analogous to Ohm's law between probability current and Information affinity—which has hitherto been discussed primarily in theoretical terms—can be realised in a real system and quantitatively verified. By electrically driving magnetic skyrmions undergoing Brownian motion without material transport, we directly measured information transfer via skyrmion–skyrmion interactions. Using real-time optical tracking and feedback control, we made one skyrmion function as Maxwell's demon whilst allowing the other skyrmion to absorb heat from the environment. By continuously adjusting the intensity of the feedback current, we demonstrated Ohm's–law–like relationship in an information circuit. We also showed that the information conductance varies due to corrections arising from scattering.

These results experimentally demonstrate that the concepts of circuit theory can be extended to information transport by considering the size of the state space and the scattering of the probability current. The findings of this study may provide a framework for analysing non-equilibrium information transport in physical, chemical and biological systems, in addition to applications in ultra-low-power information processing, such as Brownian computation.

# Method Summary

## Sample fabrication

Sample A: Si/$SiO_2$ substrate | Ta(5) | Co-Fe-B(1.00) | Ta(0.34) | MgO(1.5) | $SiO_2$(3), and Sample B: Si/$SiO_2$ substrate | Ta(5) | Co-Fe-B(1.22) | Ta(0.21) | MgO(1.5) | $SiO_2$(3) (film thickness in nm) were fabricated at 20 °C using a magnetron sputtering system (ULVAC, QAM-4-STS). In these samples, skyrmions appeared in the Ta | Co-Fe-B | Ta | MgO layered structure [27]. For the deposition of each layer, composite targets of $Co_{16}Fe_{64}B_{20}$ (where the numbers represent the atomic percentages of each element), MgO and $SiO_2$, as well as a metallic Ta target, were used for sputtering in an Ar atmosphere. The film thicknesses reported in this study are nominal values, and their absolute accuracy is limited. However, precise control of relative changes in film thickness is essential for the reproducible experiments. After film deposition, the samples were annealed on a hot plate in air at 150 °C for 3 minutes [46].

After deposition, the skyrmion film was patterned using maskless lithography (Heidelberg Ltd., MLA150/Neoark Palet) and an Ar ion milling system (Hakuto Ltd., 10IBE) to enable the application of an electric current. The gap width between the two lines is 1.5 μm. Subsequently, a 2 nm-thick $SiO_2$ layer was deposited to confine the skyrmions within the potential boxes [39]. In Sample A, the diffusion coefficient within this layer reaches approximately 60 $\mu m^2/s$. The results presented in this paper were obtained using Sample A. Sample A contains multiple devices; the results presented here are from the device that remained stable for the longest duration during measurement. Similar results were obtained for other devices, but the period during which stable measurements could be taken was shorter, resulting in a smaller data set. Similar data regarding the positive nominal feedback current were also obtained for Sample B, but it failed before tests under negative bias could be carried out.

## Measurement conditions

As shown in Figure 1c, d, skyrmions were observed using a magneto-optical Kerr effect (MOKE) microscope under a vertical bias magnetic field of approximately 0.4 mT. The size of the observed skyrmions ranged from 1 to 2 μm, as shown in Figure 1d. A DC probe was brought into contact with the upper surface of the skyrmion film, and current was injected in the in-plane direction using a Keithley 2400 source meter. Python 3.11 was used for feedback control, and a real-time tracking programme was developed using OpenCV and pyvisa. The feedback delay time was typically less than 10 ms. The camera frame rate was 100 fps, and the total number of frames exceeded 3.5 million, including over 60,000 transition events. All experiments were conducted at room temperature (T = 298.5 K). The temperature was precisely controlled using a VAHEAT system (Interference).

## Feedback method

In this experiment, the position of the system's skyrmion was continuously observed using a magneto-optical microscope, and feedback was achieved by applying a current to the wire containing the 'demon'. When the system's skyrmion was in the '1' state and the nominal feedback current was positive, a positive current was applied to the wire containing the 'demon', thereby promoting the transition of the demon from vertex I to vertex II. Conversely, when the skyrmion in the system was in the '0' state and the nominal feedback current was positive, a negative current was applied to promote the transition from vertex III to IV. If the sign of the nominal current was negative, the reverse current promoted transitions from II to I and from IV to III. Throughout this entire feedback cycle, no external force was directly applied to the skyrmion in the system.

## Analysis method

We describe the analytical method for evaluating information thermodynamic quantities. The probability current $J_n$ flowing between adjacent states $n$ and $n$+1 is described, as shown in Equation (1), using the transition rate $W_{n\to n+1}$ and the state probability $p$. As shown in Figure 2a, the four edges are denoted by $n$ = I, II, III and IV. The net probability current along the closed loop formed by these states is denoted by $J(C)$.

The sum of the information affinities applied to the edges on the system side is defined as follows.

$$F_{\mathrm{S}}^{\mathrm{Info}} \equiv F_{\mathrm{II}}^{\mathrm{Info}} + F_{\mathrm{IV}}^{\mathrm{Info}} = k_{\mathrm{B}} \ln\left(\frac{p_{\mathrm{II}}}{p_{\mathrm{III}}}\right) + k_{\mathrm{B}} \ln\left(\frac{p_{\mathrm{IV}}}{p_{\mathrm{I}}}\right) = k_{\mathrm{B}} \ln\left(\frac{p_{\mathrm{II}} p_{\mathrm{IV}}}{p_{\mathrm{III}} p_{\mathrm{I}}}\right). \tag{11}$$

The HE information flow is obtained from the product of the probability current and the information affinity, as shown in Equation (4).

Independently of the above definition based on probability current and information affinity, the HE information flow was experimentally evaluated [31] using mutual information calculated from binary data on skyrmion positions obtained by a MOKE microscope. The mutual information $I$ between two skyrmions is expressed as follows using Shannon entropy S.

$$I(x_{n+k}^{a}\colon x_{n}^{b}) = S(x_{n+k}^{a}) - S(x_{n+k}^{a}|x_{n}^{b}) = \sum_{x,y} p\,(x_{n+k}^{a}, x_{n}^{b}) \ln\left[\frac{p(x_{n+k}^{a}, x_{n}^{b})}{p(x_{n+k}^{a})p(x_{n}^{b})}\right], \tag{12}$$

Here, $a$ and $b$ = $D$ and $S$, where $D$ and $S$ represent the skyrmions of the demon and the system, respectively. The HE information flow is obtained from the time derivative of the mutual information as follows.

$$\dot{I}_{\mathrm{Sys}} = \lim_{\Delta t \to 0} \frac{I(x_{n+1}^{s}\colon x_{n}^{d}) - I(x_{n}^{s}\colon x_{n}^{d})}{\Delta t} \tag{13}$$

In the steady state, the bidirectional information flow between the system and the demon is of equal magnitude, satisfying $\dot{i}_{\mathrm{Sys}} + \dot{i}_{\mathrm{demon}} = 0$. Details of the analysis of mutual information are presented in Supplementary Information S1.

Statistical uncertainties were evaluated for all error bars using the bootstrap method. For each measurement video, resampling was performed using the reconstruction sampling method, and the relevant quantities were recalculated for each bootstrap sample. This procedure was repeated one million times for each applied current. For all error bars, a central 99.73 per cent interval, equivalent to 3σ and derived from the 0.135th and 99.865th percentiles of the block bootstrap distribution (with each video serving as a single block), was used. To account for sample degradation, with the elapsed time since the start of the feedback measurement denoted as $t$ [hours] and the measured quantity as $X$, an additional systematic uncertainty of $0.005t|X|$ was included in the direction of increasing signal amplitude. This systematic contribution was added to the statistical uncertainty.

## Acknowledgements

This research was supported by the Japan Society for the Promotion of Science (JSPS) Grant-in-Aid for Scientific Research (S) JP20H05666, Basic Research (B) 26K01383, Research Activity Start-up Support 24K22860, Young Researchers Programme JP26K1707, JST CREST JPMJCR20C1, the Ministry of Education, Culture, Sports, Science and Technology's 'Next-Generation X-nics Semiconductor Creation Hub Formation Project (X-NICS)' JPJ011438, and the Asahi Glass Foundation. A part of this work was supported by "Advanced Research Infrastructure for Materials and Nanotechnology in Japan (ARIM)" of the Ministry of Education, Culture, Sports, Science and Technology (MEXT) (Proposal Number JPMXP1226AT0105).

## Author contributions

S. Miki, Y. Suzuki, and R. I. conceived and designed the research project. S. Miki Y. Suzuki, and K. A. also carried out the measurement and analysis of the experimental data. S. Miki developed the experimental apparatus, including the measurement programme. R. I., Y. H., S. Morishita, Y. Shiota, and T. N. fabricated the samples and devices. H. I. and E. T. developed the simulation code, whilst S. Miki and H. N. performed the simulations. M. G., Y. Suzuki, and S. Mizukami guided and supervised this research.

# Author information

The authors declare that they have no competing financial interests. Correspondence and requests for materials should be addressed to Miki (soma.miki.d8@tohoku.ac.jp).